\documentclass[11pt,twoside]{article}
\usepackage{atmp}
\copyrightnotice{2003}{7}{381}{405}
\usepackage{latexsym}
\usepackage[all]{xy}

\newcommand{\R}{\mathbf{R}}
\newcommand{\sHom}{\underline{\mathrm{Hom}}}
\newcommand{\ra}{\rightarrow}

\newcommand{\iso}{\cong}
\newcommand{\Ld}{\mathbf{L}}
\newcommand{\lotimes}{\stackrel{\Ld}{\otimes}}
\newcommand{\D}{{\mathbf D}_{\mathrm{coh}}^b}
\newcommand{\scdot}{\,\cdot\,}

\theoremstyle{plain}
\newtheorem{theorem}{Theorem}[section]

\newtheorem{proposition}[theorem]{Proposition}

\newcommand{\cG}{ {\cal G} }
\newcommand{\cF}{ {\cal F} }

\newcommand{\cI}{ {\cal I} }

\newcommand{\Hom}{ \mbox{Hom} }
\newcommand{\Ext}{ \mbox{Ext} }

\begin{document}

\title{D-branes, $B$ fields, and Ext groups}

\url{hep-th/0302099}

\author{Andrei C\u ald\u araru$^1$, Sheldon Katz$^2$, Eric Sharpe$^2$}
\address{$^1$Department of Mathematics \\
University of Pennsylvania \\
David Rittenhouse Lab.\\
209 South 33rd Street \\
Philadelphia, PA  19104-6395 }

\address{$^2$Department of Mathematics \\
1409 W. Green St., MC-382 \\
University of Illinois \\
Urbana, IL  61801}

\addressemail{andreic@math.upenn.edu}
\addressemail{katz@math.uiuc.edu}
\addressemail{ersharpe@uiuc.edu}

\markboth{\it D-BRANES, $B$ FIELDS, AND EXT GROUPS}{\it A. C\u
  ALD\u ARARU, S. KATZ, E. SHARPE}

\begin{abstract}
In this paper we extend previous work on calculating
massless boundary Ramond sector spectra of open strings
to include cases with nonzero flat $B$ fields.
In such cases, D-branes are no longer well-modeled precisely
by sheaves, but rather they are replaced by `twisted' sheaves,
reflecting the fact that gauge transformations of the $B$ field
act as affine translations of the Chan-Paton factors.
As in previous work, we find that the massless boundary Ramond
sector states are counted by Ext groups -- this time,
Ext groups of twisted sheaves.  As before, the computation of
BRST cohomology relies on physically realizing some spectral
sequences.  Subtleties that cropped up in previous work also appear
here.
\end{abstract}

\newpage

\section{Introduction}

One of the predictions of current proposals for the physical
significance of derived categories (see \cite{paulron,medc,mikedc,paulalb}
for an incomplete list of early references) is that
massless boundary Ramond sector states of open strings
connecting D-branes wrapped on complex submanifolds
of Calabi-Yau's, with holomorphic gauge bundles, should
be counted by mathematical objects known as Ext groups.
Specifically, if we have D-branes wrapped on the
complex submanifolds $i: S \hookrightarrow X$ and
$j: T \hookrightarrow X$ of a Calabi-Yau $X$,
with holomorphic gauge fields in\footnote{Because of the Freed-Witten
anomaly, gauge fields on D-brane worldvolumes do not couple to honest
bundles, but rather to twisted bundles, which is reflected
in the $\sqrt{K_S^{\vee}}$ factors.  A detailed explanation
of the sheaf-brane correspondence, taking into account the Freed-Witten
anomaly, appears in \cite{orig}.} ${\cal E} \otimes
\sqrt{K_S^{\vee}}$, ${\cal F}\otimes\sqrt{K_T^{\vee}}$,
respectively, then massless boundary Ramond sector states
should be in one-to-one correspondence with elements of
either
\begin{displaymath}
\mbox{Ext}^*_X\left( i_* {\cal E}, j_* {\cal F} \right)
\end{displaymath}
or
\begin{displaymath}
\mbox{Ext}^*_X\left( j_* {\cal F}, i_* {\cal E} \right)
\end{displaymath}
(depending upon the open string orientation).

In \cite{orig} we explicitly checked this proposal
for large-radius Calabi-Yau's,
by using standard well-known BCFT methods to compute 
the massless Ramond sector spectrum of open strings connecting
such D-branes, and then relating that spectrum to Ext groups.
Naively the massless Ramond sector states are not counted
by Ext groups, but rather sheaf cohomology groups that
are merely related to Ext groups via spectral sequences.
However, a closer examination of the physics revealed
that, at least in cases we understand well,
those spectral sequences are realized
directly in BRST cohomology, and so the massless Ramond sector
spectrum is counted directly by Ext groups.

In \cite{kps} we extended the work in \cite{orig} to consider
D-branes in orbifolds.  We again found that massless boundary
Ramond sector states are counted by Ext groups, this time
on quotient stacks.  Previously, open string spectra in orbifolds
had only been computed for topologically trivial brane configurations,
so we were able to implicitly solve the old technical challenge
of calculating spectra for more general configurations.
We also commented on the physical meaningfulness of quotient stacks.
For example, fractional branes cannot be described as sheaves
on quotient spaces -- they can only be described as sheaves on quotient
stacks.   

Both of the papers \cite{orig,kps} implicitly assumed that there
was no $B$ field background.
In this paper we shall extend the methods of \cite{orig}
to the case of nonzero, nontrivial, flat $B$ fields.

In the presence of a nontrivial $B$ field,
one does not have an honest bundle on D-brane worldvolumes,
essentially
because in open strings gauge transformations of the $B$ field
mix with transformations of the Chan-Paton factors, as
\begin{eqnarray*}
B & \mapsto & B \: + \: d \Lambda \\
A & \mapsto & A - \Lambda I
\end{eqnarray*} 
where $I$ is the identity matrix.
Instead, the bundle is
twisted, in the sense that the transition functions $g_{\alpha \beta}$
no longer close to the identity on triple overlaps, but rather
obey
\begin{displaymath}
g_{i j} \, g_{j k} \, g_{k i} \: = \:
\alpha_{i j k}
\end{displaymath}
for some $U(1)$-valued 2-cocycle $\alpha$ partly defining the $B$ field.
(For a readable description of nontrivial $B$ fields, see \cite{hitchin}).
This twisting has played an important role in, for example,
the derivation \cite{medt,dtrev}
of Douglas's description of discrete torsion for D-branes \cite{dougdt},
and has also been discussed in \cite{kap,bm}.

One can make rigorous mathematical sense of such twisted
bundles \cite{andreithesis}, as well as analogous twisted notions
of sheaf cohomology and Ext groups, as we shall review.

In this paper, after discussing how nontrivial flat $B$ fields
can be described by certain elements of the group
$H^2(X, {\cal O}_X^*)$, we shall compute the massless boundary
Ramond sector spectrum of open strings connecting D-branes,
in the presence of a $B$ field defined by $\omega \in
H^2(X, {\cal O}_X^*)$, and we shall show that such states are
in one-to-one correspondence with elements of twisted
Ext groups, either
\begin{displaymath}
\mbox{Ext}^*_{X, \omega}\left( i_* {\cal E}, j_* {\cal F} \right)
\end{displaymath}
or
\begin{displaymath}
\mbox{Ext}^*_{X, \omega}\left( j_* {\cal F}, i_* {\cal E} \right)
\end{displaymath}
(depending upon the open string orientation).

We begin in section~\ref{overview} with an overview of bundles
and sheaves twisted by the $B$ field, and relevant bits of physics.
In section~\ref{parcoin} we discuss massless boundary Ramond sector
spectra for open strings between parallel coincident branes.
The analysis proceeds very much as in \cite{orig,kps}, and we find
that said spectra are ultimately counted by Ext groups between
twisted sheaves.
In section~\ref{ex} we discuss an example.  Unfortunately,
due to a lack of known Calabi-Yau manifolds with the relevant properties,
we are not able to discuss a wide variety of examples, but we do
at least illustrate how the relevant calculations are performed.
In section~\ref{pardiff} we discuss massless boundary Ramond sector
spectra for parallel branes on submanifolds of different dimension.
In section~\ref{gen} we discuss massless boundary Ramond sector spectra
for general branes.  Finally, in appendix~\ref{pfs} we prove some
spectral sequences used in this paper.

\section{Overview of twisted sheaves}   \label{overview}

As mentioned in the introduction, in the presence
of a nontrivial $B$ field, one no longer has an
honest bundle on the worldvolume of a D-brane.
Ordinarily on triple overlaps the transition functions $g_{i j}$
of a bundle obey the condition
\begin{displaymath}
g_{i j} g_{j k} g_{k i}
\: = \: 1
\end{displaymath}
However, $B$ field gauge transformations combine nontrivially
with Chan-Paton translations, as
\begin{eqnarray*}
B & \mapsto & B \: - \: d \Lambda \\
A & \mapsto & A \: + \: \Lambda I
\end{eqnarray*}
so if the $B$ field has nonzero transition functions,
then the connection on our `bundle' picks up affine translations
between coordinate charts.
The effect is that the transition functions $g_{i j}$
of the `bundle' obey the condition \cite{freeded}
\begin{equation}
\label{cochain}
g_{i j} g_{j k} g_{k i}
\: = \: \alpha_{i j k} 
\end{equation} 
for a 2-cocycle $(\alpha_{i j k})$ representing a cohomology class
in $H^2(X, C^{\infty}(U(1)))$.  These cohomology classes define gerbes,
and the $\{g_{i j}\}$ satisfying (\ref{cochain}) define twisted bundles.
{}From the exact sequence
\[
0\to \mathbf{Z}\to C^{\infty}(\mathbf{R}) \to
C^{\infty}(U(1)) \to 0
\]
we get a map $H^2(X,C^{\infty}(U(1)))\to H^3(X,\mathbf{Z})$ which 
takes $[\alpha]$ to a class in $H^3(X,\mathbf{Z})$, the characteristic
class of the gerbe.  Note that
\begin{displaymath}
H^2(X, C^{\infty}(U(1))) \cong H^3(X, {\bf Z}).
\end{displaymath}
See \cite{hitchin} for more details.

Note that modulo scalars, (\ref{cochain})
becomes
\begin{equation}
\label{azumaya}
\bar{g}_{i j} \bar{g}_{j k} \bar{g}_{k i}
\: = \: 1
\end{equation}
where $\bar{g}$ is the $\mathrm{PGL}(n)$ image of $g \in \mathrm{GL}(n)$.
If the twisted bundle has rank $r$, then (\ref{azumaya}) says that we
get a cohomology class $[\bar{g}]\in H^1(X, C^{\infty}(\mathrm{PGL}(r)))$.
{}From the exact sequence
\[
0\to \mathbf{Z}_r\to C^{\infty}(\mathrm{GL}(r)) \to
C^{\infty}(\mathrm{PGL}(r)) \to 0
\]
we get a coboundary map $H^1(X, C^{\infty}(\mathrm{PGL}(r)))\to 
H^2(X,\mathbf{Z}_r)$, 
taking $[\overline{g}]$ to an element of $H^2(X,\mathbf{Z}_r)$.  It is 
straightforward to check that the image of this class in 
$H^2(X, C^{\infty}(U(1)))$ induced by the inclusion $\mathbf{Z}_r\to U(1)$
is precisely the class of the gerbe defined by
the original $\alpha$.  Since $H^2(X,\mathbf{Z}_r)$ is torsion, we see that
rank $r$ twisted bundles can only exist if the class of the underlying gerbe
is $r$ torsion. In particular, if the gerbe is not a torsion class, then there
are no finite-rank twisted bundles at all.

Such twisted bundles are understood in mathematics
(see for example \cite{andreithesis}).
One can define not only twisted bundles, but also twisted
coherent sheaves on a space.
In fact, sheaves can also be defined in terms of local data
and transition functions obeying a cocycle condition
(see for example \cite[cor. I-14]{eh}), so we can define twisted
sheaves in exactly the same form as twisted bundles.
On the other hand, although twisted bundles are immediately
realized in D-brane physics, the physics of twisted sheaves
is more obscure.  Part of what we are doing in this paper
amounts to giving concrete evidence that D-branes in flat
$B$ field backgrounds can be accurately modeled by twisted sheaves
(see also \cite{kaporlov} for previous work in this vein).

In this holomorphic context, the twisting is defined by
an element $\alpha\in H^2(X, {\cal O}_X^*)$ associated
to the $B$ field.  Moreover, one can also define Ext groups
between twisted sheaves.  If ${\cal S}$, ${\cal T}$ are two
coherent sheaves on $X$, both twisted by the same
element $\omega \in H^2(X, {\cal O}_X^*)$, then one can
define $\mbox{Ext}^*_{X, \omega}\left( {\cal S}, {\cal T} \right)$.
We will say more about such Ext groups and how to arrive at
elements of $H^2(X, {\cal O}_X^*)$ from a $B$ field momentarily.
First we will review twisted sheaves in more detail.

For convenience of the reader, we give the complete definition of a 
twisted sheaf.  If $\alpha\in H^2(X,
{\cal O}_X^*)$ is represented by a \v{C}ech 2-cocycle 
$(\alpha_{ijk})$ on an
open cover $\{U_i\}_{i\in I}$, with the $\alpha_{ijk}$ holomorphic,
an $\alpha$-twisted sheaf $\cF$ consists of a
pair 
\[ (\{\cF_i\}_{i\in I}, \{\phi_{ij}\}_{i,j\in I}), \]
where $\cF_i$ is a sheaf on $U_i$ for all $i\in I$ and
\[ \phi_{ij}: \cF_j|_{U_i\cap U_j} \ra \cF_i|_{U_i\cap U_j} \]
is an isomorphism for all $i,j\in I$, subject to the conditions:
\begin{enumerate}
\item $\phi_{ii} = \mathrm{Id}$;
\item $\phi_{ij} = \phi_{ji}^{-1}$;
\item $\phi_{ij}\circ \phi_{jk}\circ \phi_{ki} = \alpha_{ijk} 
\cdot \mathrm{Id}$.
\end{enumerate}

The class of twisted sheaves together with the obvious notion of
homomorphism is an abelian category,
the category of $\alpha$-twisted sheaves, leading to the stated
notion of Ext groups by general homological algebra. 

This notation is consistent, since one can prove that this category
is independent of the choice of the covering
$\{U_i\}$~(\cite[1.2.3]{andreithesis}) or of the particular cocycle
$\{\alpha_{ijk}\}$~(\cite[1.2.8]{andreithesis})

Now a $B$-field is a connection on a gerbe.  To say it is a
connection means that we can express it locally as a 2-form $B_i$,
subject to gauge transformations $B_i-B_j=d\Lambda_{ij}$ for some
1-forms $\Lambda_{ij}$.  The field strength $H=dB$ is then globally
defined.  Since $\Lambda_{ij}+\Lambda_{jk}+\Lambda_{ki}$ is exact, we
can locally write
$\Lambda_{ij}+\Lambda_{jk}+\Lambda_{ki}=-i\alpha_{ijk}^{-1}d
\alpha_{ijk}$, where $\alpha_{ijk}$ are local $U(1)$-valued functions,
once again representing the class of the underlying gerbe. In this
context, $[H]\in H^3(X,\mathbf{R})$ is the characteristic class of the
gerbe (it actually lies in $H^3(X,\mathbf{Z})$).  If in addition the
$B$ field is flat, i.e.\ if $H=0$, then the above considerations lead
to the holonomy class of $B$ in $H^2(X,\mathbf{R}/\mathbf{Z})$.  See
\cite{hitchin} for more details.  In particular the contribution of the
$B$-field to the action can be specified topologically for a closed
string worldsheet.

A connection on a twisted $U(N)$ bundle in the presence of a $B$-field is 
analogous to the usual notion of a connection.  We have a local collection
of 1 forms $A_i$ satisfying
\[
A_i\: -\: \left(f_{ij} \right) A_j\left( f_{ij} \right)^{-1} \:
= \: f_{ij}^{-1}df_{ij}\: -\: I \Lambda_{ij}.
\]
Here, $\Lambda_{ij}$ is the gauge transformation of the $B$-field as discussed
above, and $I$ is the $N \times N$ identity matrix.  
Note that $\mbox{Tr }F$ is not gauge invariant, but $\mbox{Tr }F+B$ is gauge invariant.

For completeness, we describe the
relationship between elements of the group $H^2(X, {\cal O}_X^*)$
and gerbes, at least in the case of a Calabi-Yau threefold.  Extensions
are straightforward.
There is an exact sequence
\begin{displaymath}
H^2(X, {\cal O}_X ) \: \longrightarrow \:
H^2(X, {\cal O}_X^* ) \: \longrightarrow \:
H^3(X, {\bf Z} ) \: \longrightarrow \: H^3(X, {\cal O}_X )
\end{displaymath}
and on a Calabi-Yau threefold, $H^2(X, {\cal O}_X) = 0$.  Thus $H^2(X,
{\cal O}_X^* )$ is the kernel of 
\begin{displaymath}
H^3(X, {\bf Z} ) \to H^3(X, {\cal
O}_X )\simeq H^{0,3}(X).
\end{displaymath}
This map can be described by first mapping
$H^3(X, {\bf Z} )\to H^3(X, {\bf R} )$, then following with the
Hodge-theoretic projection of a de Rham third cohomology class onto
its $(0,3)$ component.  Since an element of $H^3(X, {\bf Z} )$ is
real, the $(3,0)$ component vanishes whenever its $(0,3)$ component
vanishes.  Thus we see that given an element of $H^3(X,{\bf Z})\cap
(H^{2,1}(X)\oplus H^{1,2}(X))$,\footnote{We are not being quite precise in
notation here,
as $H^3(X, {\bf Z})$ does not embed in $H^3(X, {\bf R})$ in the presence of
torsion.} we can uniquely reconstruct a
corresponding element of $H^2(X, {\cal O}_X^*)$.  As an aside,
we should mention that the group
$H^3(X,{\bf Z})\cap (H^{2,1}(X)\oplus H^{1,2}(X))$ plays an important role
in the generalized Hodge conjecture in dimension three; see appendix~\ref{ghc}
for a few more details.

In this paper, we are only interested in flat\footnote{Non-flat $B$ fields
typically break supersymmetry in Calabi-Yau compactifications.  In 
non-Calabi-Yau
compactifications where they are used, non-flat $B$ fields
typically fix K\"ahler moduli,
usually preventing one from going to weak-coupling limits where the analysis of
this paper is relevant.  For these reasons, we shall not consider
non-flat $B$ fields in this paper.  We should also mention,
though, that the mathematical analysis
of appendix~A does carry over to the case of non-flat $B$ fields when the
D-branes are wrapped on a submanifold such that the restriction of the
$B$ field curvature to the submanifold is torsion.} $B$ fields,
which means that the curvature (the de Rham image of the
characteristic class) vanishes, so the corresponding
element of  $H^3(X, {\bf Z})$ is purely torsion.  
When this happens, its
image in $H^3(X, {\cal O}_X )$ necessarily vanishes for the above reasons.

In summary, the torsion subgroup of $H^2(X,{\cal O}_X^*)$ is isomorphic to the
torsion subgroup of $H^3(X,{\bf Z})$.  So the elements 
of $H^2(X,{\cal O}_X^*)$ relevant to our study of D-branes correspond to 
purely topological information.

As we have seen above, one can define holomorphic {\it locally free}
twisted sheaves of finite rank only when the twisting
is by an element of $H^2(X, {\cal O}_X^*)$ corresponding
to a torsion element of $H^3(X,{\bf Z})$.  The elements arising this way
form the {\em Brauer group}.
(A widely-believed conjecture of Grothendieck says that the elements 
of the Brauer group
are the same as the torsion elements of $H^2(X, {\cal O}_X^*)$.)
For twistings by non-torsion elements, no locally free twisted
sheaves of finite rank exist. 
So our physical constraint of flat $B$ fields translates
directly into mathematics.

In passing, we will note that 
there is a closely analogous phenomenon in the study of WZW models.
There, one has strings propagating on a group manifold with a nontrivial
$B$ field background, in which the curvature $H$ of the $B$ field
is nonzero (this is the point of the Wess-Zumino term, after all).
D-branes cannot wrap the entire group manifold.  Instead,
D-branes can only wrap submanifolds $S$ such that the restriction
of the characteristic class to $S$ is torsion \cite{wzw1,wzw2}, mirroring the
statement that finite-rank locally-free twisted sheaves only exist when
the characteristic class of the twisting is torsion.

Another way to think about twisted holomorphic bundles is that they
are associated to projective space bundles that cannot be obtained
by projectivizing an ordinary holomorphic vector bundle.
In detail, given a projective space bundle, imagine lifting 
the fibers locally from ${\bf P}^n$ to ${\bf C}^{n+1}$.
The original projective space bundle was a {\it bundle},
so its transition functions close on triple overlaps.
However, if we lift those same transition functions from
${\bf P}^n$ to ${\bf C}^{n+1}$, then they need only close
up to the action of ${\bf C}^{\times}$ on triple overlaps.
In other words, given a projective space bundle, it need not
lift to an honest vector bundle, but in general will only
lift to a twisted vector bundle, twisted by an element of
$H^2\left(X, {\cal O}_X^*\right)$ defined by the failure of the
transition functions to close on triple overlaps.

Given two twisted sheaves ${\cal E}$, ${\cal F}$, one can
define a twisted sheaf of local homomorphisms $\underline{
\mbox{Hom}}({\cal E}, {\cal F} )$.
If ${\cal E}$ is twisted by $\omega \in H^2(X, {\cal O}_X^*)$
and ${\cal F}$ is twisted by $\omega' \in H^2(X, {\cal O}_X^*)$,
then $\underline{\mbox{Hom}}({\cal E}, {\cal F})$ is
a twisted sheaf, twisted by $\omega^{-1} \cdot \omega' \in
H^2(X, {\cal O}_X^*)$.

In particular, if two bundles ${\cal E}$, ${\cal F}$ are twisted
by the {\it same} element of $H^2(X, {\cal O}_X^*)$,
then $\underline{\mbox{Hom}}({\cal E}, {\cal F})$ is an
ordinary, untwisted sheaf.

We can use that fact to set up homological algebra in the
usual form.  If ${\cal S}$, ${\cal T}$ are two twisted sheaves,
but twisted by the {\it same} element of $H^2(X, {\cal O}_X^*)$,
then we can define local $\underline{\mbox{Ext}}^n_{ {\cal O}_X }
\left( {\cal S}, {\cal T}\right)$ and global $\mbox{Ext}^n_X
\left( {\cal S}, {\cal T} \right)$.

This last fact is directly relevant for this paper,
as in any given theory there is only a single $B$ field that
twists the D-brane bundles.  Thus, open string spectra are
always computed between twisted sheaves that have been
twisted by the same element of $H^2(X, {\cal O}_X^*)$,
which means that for all our string spectrum calculations,
there exists a relevant notion of Ext groups
in mathematics, and the associated Ext sheaves are ordinary sheaves.

Serre duality functions for twisted sheaves much as for 
ordinary untwisted sheaves.
For $\omega \in H^2(X, {\cal O}_X^*)$ an element of the Brauer
group ({\it i.e.} corresponding to a flat $B$ field),
for any two $\omega$-twisted coherent sheaves ${\cal S}$,
${\cal T}$ on a Calabi-Yau manifold $X$, we have
\begin{displaymath}
\mbox{Ext}^i_{X, \omega}\left( {\cal S}, {\cal T} \right) \: \cong \:
\mbox{Ext}^{n-i}_{X, \omega}\left( {\cal T}, {\cal S} \right)^*
\end{displaymath}

We have spoken a great deal so far about twisted sheaves
in general terms.  Although our remarks apply in principle,
in practice we unfortunately do not know of any examples
of smooth Calabi-Yau manifolds with non-zero torsion part of
$H^2(X, {\cal O}_X^*)$.

We do know of some spaces that are close but not quite Calabi-Yau,
and for the purposes of illustrating our methods with some
concrete examples, we will mention a few such examples.

Our example of a space with nonzero torsion in $H^2(X, {\cal O}_X^*)$ is taken
directly from the appendix of
\cite{agm}.  
Let $W \subset {\bf P}^3 \times {\bf P}^3$ be a generic hypersurface
of bidegree (2,2), and let $K \subset {\bf P}^3$ be the discriminant
locus of the fibration $p_1: W \rightarrow {\bf P}^3$, where
$p_1$ is the projection onto the first factor.  One can show that
$K$ is an octic surface with 80 ordinary nodes.  Let $d: X \rightarrow
{\bf P}^3$ be the double cover of ${\bf P}^3$ branched over $K$,
and let $\pi: \tilde{X} \rightarrow X$ be the blowup of the
80 nodes of $X$, so that $\tilde{X}$ is smooth.
Since we have blown-up the nodes, rather than small-resolved them,
this space is not quite Calabi-Yau, but in some sense, close.
As shown in \cite{agm}, this space has 2-torsion in $H^2(X, {\cal O}_X^*)$.

\section{Parallel coincident branes on $S \hookrightarrow X$ }  \label{parcoin}

Let $i: S \hookrightarrow X$ be a smooth complex submanifold of
a Calabi-Yau $X$, and let $\omega \in H^2(X, {\cal O}_X^*)$ define
a flat $B$ field.  Let ${\cal E}$, ${\cal F}$ be two $i^* \omega$-twisted
holomorphic bundles on $S$, corresponding to the twisted bundles
on the worldvolumes of two sets of branes, both wrapped on $S$.

Since we are assuming the $B$ field is flat, locally we can
gauge the $B$ field to zero, so the physics analysis proceeds exactly as 
in \cite{orig}, except that the Chan-Paton factors now couple to
twisted bundles rather than honest bundles \cite{freeded}.\footnote{Note
that we can sometimes make a similar gauge choice even if the $B$-field is not
flat, namely if we only assume that the $B$ field becomes flat after 
restriction to $S$.  So our subsequent physical analysis can proceed somewhat
more generally than we consider in this paper.  Mathematically, we can 
replace the hypothesis that the twisting 
class $\alpha\in H^2(X, {\cal O}_X^*)$ is torsion with the weaker hypothesis
that its restriction $\alpha|_S\in H^2(S,{\cal O}_S^*)$ is torsion.  In 
this situation $\alpha_S$-twisted bundles on $S$ exist.  The 
mathematical proofs in Appendix~\ref{pfs} can be extended to 
apply in this more general situation.}
We shall omit the detailed explanation (consult \cite{orig} for details).
The boundary Ramond sector states are of the form
\begin{displaymath}
b^{\alpha \beta j_1 \cdots j_m}_{\overline{\imath}_1 \cdots 
\overline{\imath}_n}(\phi_0) \, \eta^{ \overline{\imath}_1} \cdots
\eta^{ \overline{\imath}_n} \theta_{j_1} \cdots \theta_{j_m}
\end{displaymath}
where
\begin{eqnarray*}
\eta^{\overline{\imath}} & = & \psi_+^{\overline{\imath}} 
\: + \:
\psi_-^{\overline{\imath}} \\
\theta_i & = & g_{i \overline{\jmath}} \left(
\psi_+^{\overline{\jmath}} \: - \: \psi_-^{\overline{\jmath}} \right)
\end{eqnarray*}
as BRST invariants in the standard notation for the $B$ model \cite{edtft},
subject to the boundary conditions $\eta^{\overline{\imath}} = 0$
for Dirichlet directions, and {\it naively} $\theta_i = 0$ for
Neumann directions.  (The boundary conditions are modified
by the Chan-Paton factors, as in \cite{abooetal,aboo2}, a point we shall
return to momentarily.)
The indices $\alpha$, $\beta$ are Chan-Paton indices, describing
the $i^* \omega$-twisted holomorphic bundles ${\cal E}$, ${\cal F}$ on $S$.

Naively, physical states of the form above should be in one-to-one
correspondence with elements of the sheaf cohomology groups
$H^n(S, {\cal E}^{\vee} \otimes {\cal F} \otimes \Lambda^m {\cal N}_{S/X})$.

Note that although ${\cal E}$ and ${\cal F}$ are both twisted bundles,
instead of honest bundles, we can still speak of ordinary bundle-valued
sheaf cohomology, since the product ${\cal E}^{\vee} \otimes {\cal F}$
is untwisted.

As in \cite{orig},
there is a spectral sequence
\begin{displaymath}
H^p\left(S, {\cal E}^{\vee} \otimes {\cal F} \otimes \Lambda^q {\cal N}_{S/X}
\right) \: \Longrightarrow \:
\mbox{Ext}^{p+q}_{ X, \omega }\left( i_* {\cal E}, i_* {\cal F} \right)
\end{displaymath}
relating the sheaf cohomology above to the desired ($\omega$-twisted)
Ext groups.  (See appendix~\ref{pfs} for proofs.)

How is this spectral sequence realized physically?
As in \cite{orig}, there are two important subtleties
that must be taken into account:
\begin{enumerate}
\item $TX|_S$ need not split holomorphically, {\it i.e.} $TX|_S
\not\cong TS \oplus {\cal N}_{S/X}$ in general, so it is not quite right
to interpret the vertex operators as ${\cal N}_{S/X}$-valued forms.
\item The boundary conditions are twisted, by the curvature of the
Chan-Paton factors and the $B$ field (see \cite{abooetal}, 
\cite[(3.3)]{aboo2}),
so that for Neumann directions,
$\theta_i \neq 0$, but rather $\theta_i = \left( \mbox{Tr } 
F_{i \overline{\jmath}}+B_{i \overline{\jmath}} \right)
\eta^{ \overline{\jmath} }$.
\end{enumerate}
Just as in \cite{orig}, in the special case that ${\cal E} = {\cal F}$,
we shall see explicitly that taking care of these two subtleties has the
effect of physically realizing the spectral sequence in BRST cohomology.
Furthermore, we conjecture that the same result is true in general:
being careful about these subtleties should have the effect of realizing
all spectral sequences we discuss in this paper directly in BRST cohomology,
so that the physical massless boundary Ramond sector states are in 
one-to-one correspondence with
elements of Ext groups.

Let us specialize to the case ${\cal E} = {\cal F}$, to see how
the spectral sequence is realized explicitly.
As the argument is essentially identical to that in \cite{orig},
we shall be brief.
We deal with both of the subtleties above in the same form as in \cite{orig}.
The first subtlety is dealt with by lifting ${\cal N}_{S/X}$-valued forms to
$TX|_S$-valued forms, which can be directly associated to vertex operators,
unlike ${\cal N}_{S/X}$-valued forms.  The resulting $TX|_S$-valued forms need
not be holomorphic, however.  Their images under $\overline{\partial}$
are $TS$-valued forms, to which we can apply the boundary condition
$\theta_i = \left( \mbox{Tr } F_{i \overline{\jmath}}
+ B_{i \overline{\jmath}} \right)
\eta^{ \overline{\jmath} }$.  Thus, the BRST operator acting on
the $TX|_S$-valued forms can be interpreted as a composition of
$\overline{\partial}$ and an evaluation map.
For example, starting from an element of $H^0(S, {\cal E}^{\vee}
\otimes {\cal E} \otimes {\cal N}_{S/X})$, we have the composition
\begin{displaymath}
H^0(S, {\cal E}^{\vee} \otimes {\cal E} \otimes {\cal N}_{S/X} )
\: \stackrel{ \delta }{\longrightarrow} \: 
H^1(S, {\cal E}^{\vee} \otimes {\cal E} \otimes TS )
\: \stackrel{eval}{\longrightarrow} \:
H^2(S, {\cal E}^{\vee} \otimes {\cal E} )
\end{displaymath}
where the first map ($\delta$) is the coboundary map,
and the second map ($eval$) is the evaluation map realizing the
boundary conditions.
This argument is identical to that used in \cite{orig},
except for the fact that here ${\cal E}$ is an $i^* \omega$-twisted
holomorphic bundle, not an honest holomorphic bundle.
Also just as in \cite{orig}, the composition above is exactly
the differential $d_2$ of the spectral sequence.  

Finally, as in 
\cite{kps}, we still have an Atiyah class 
$a({\cal E}) \in H^{1}(S,\Omega^1\otimes
End({\cal E}))$ corresponding to the canonical extension
\begin{equation}
\label{atiyah}
0\to \Omega^1\otimes {\cal E}\to j({\cal E})\to {\cal E}\to 0
\end{equation}
of twisted sheaves defined by the twisted sheaf of 1-jets of ${\cal E}$.
The second map is simply cup product
with $a({\cal E})$.  Note that $\Omega^1\otimes
End({\cal E})$ is an ordinary sheaf.  The computation proceeds exactly
as in \cite{orig,kps}.  The only change is that the boundary conditions
get modified as we have discussed above, corresponding to
(\ref{atiyah}) being an extension of twisted sheaves rather than of 
ordinary sheaves.

Thus, we have seen explicitly that in the special case
${\cal E} = {\cal F}$, the spectral sequence is realized directly
in BRST cohomology.  We conjecture that the same statement is true
for more general ${\cal E}$, ${\cal F}$.

\section{Example}  \label{ex}

In this section we shall briefly consider an example
involving branes wrapped on a submanifold of a space with
nontrivial torsion in $H^2(X, {\cal O}_X^*)$.

Now, as mentioned earlier, we do not have any examples of
Calabi-Yau's with such a property.
However, we do have some examples of spaces that are not quite
Calabi-Yau, and for the purposes of illustrating how to perform
computations using the technology presented here, such an
example will do.  Of course, since the space $X$ is not Calabi-Yau,
our computations will be purely formal in nature -- they cannot
reflect any physics, since the B model cannot be defined on $X$.

Recall earlier we mentioned an example taken from the appendix
of \cite{agm}.  
Let $W \subset {\bf P}^3 \times {\bf P}^3$ be a generic hypersurface
of bidegree (2,2), and let $K \subset {\bf P}^3$ be the discriminant
locus of the fibration $p_1: W \rightarrow {\bf P}^3$, where
$p_1$ is the projection onto the first factor.  One can show that
$K$ is an octic surface with 80 ordinary nodes.  Let $d: X \rightarrow
{\bf P}^3$ be the double cover of ${\bf P}^3$ branched over $K$,
and let $\pi: \tilde{X} \rightarrow X$ be the blowup of the
80 nodes of $X$, so that $\tilde{X}$ is smooth.

The discussion in \cite{agm} identifies a 2-torsion element $\alpha\in
H^2(\tilde{X},\mathcal{O}_{\tilde{X}}^*)$
by exhibiting a $\mathbf{P}^1$ bundle over $X$ minus its nodes, and showing
that this is not the projectivization of a vector bundle $\mathcal{E}$.  After 
blowing up the nodes, this argument can be transferred to $\tilde{X}$.

In other words, the $\mathbf{P}^1$ bundle of \cite{agm} is the projectivization
of an $\alpha$-twisted bundle $\mathcal{E}$ on $\tilde{X}$.  

Let us imagine (formally) working with the B model on $\tilde{X}$,
and wrapping D-branes on all of $\tilde{X}$ with twisted bundle $\mathcal{E}$.
Since we are wrapping all of $\tilde{X}$, the spectral sequence is 
unnecessary.  In any case, the spectral
sequence degenerates and $\mbox{Ext}^i_{\tilde{X},\omega}
(\mathcal{E},\mathcal{E})\simeq H^i(\mathcal{E}^*\otimes\mathcal{E})$.
Note that $\mathcal{E}^*\otimes \mathcal{E}$ is an ordinary bundle (the 
associated Azumaya algebra of Brauer group theory).
In passing, note that it is only because the $B$ field is nontrivial
that we could consider such twisted bundles on the brane worldvolume -- 
were the $B$ field to be zero, for example, no such branes would exist.

Now let $D_i$ be the
exceptional divisor of the blowup of one of the 80 nodes.  
Each $D_i\simeq\mathbf{P}^1\times\mathbf{P}^1$.  In particular,
there is no torsion in $H^2(D_i,\mathcal{O}_{D_i}^*)$ or
$H^3(D_i,\mathbf{Z})$ and so the $B$ field can be trivialized after
restricting to $D_i$.  Thus the gauge fields on $D$-branes wrapping
$D_i$ live in ordinary bundles.

If $i$ and $j$ are distinct, they are disjoint and so 
\[
\mbox{Ext}^i(i_*\mathcal{O}_{D_i},j_*\mathcal{O}_{D_j})=0,
\]
where $i:D_i\hookrightarrow\tilde{X}$ and 
$j:D_j\hookrightarrow\tilde{X}$ are the natural inclusions.

Furthermore, the spectral sequence for 
\[
\mbox{Ext}^i(i_*\mathcal{O}_{D_i},i_*\mathcal{O}_{D_i})
\]
is identical to the corresponding spectral sequence in the absence of a 
$B$-field, so the presence of the $B$-field does not affect the spectrum.

\section{Parallel branes on submanifolds of different dimension}
\label{pardiff}

Let $\omega \in H^2(X, {\cal O}_X^*)$ define the $B$ field,
let $i: S \hookrightarrow X$ and $j: T \hookrightarrow X$
be smooth complex submanifolds of $X$ with $T \subseteq S$,
and let ${\cal E}$, ${\cal F}$ be $i^* \omega$-twisted holomorphic
bundles on $S$, $T$, respectively, corresponding to branes wrapped
on $S$ and $T$.

The physical analysis in this case proceeds just as in \cite{orig}.
The massless boundary Ramond sector states are of the form
\begin{displaymath}
b^{\alpha \beta j_1 \cdots j_m}_{ 
\overline{\imath}_1 \cdots \overline{\imath}_n}(\phi_0)
\eta^{ \overline{\imath}_1 } \cdots
\eta^{ \overline{\imath}_n }
\theta_{j_1} \cdots \theta_{j_m}
\end{displaymath}
where $\alpha$, $\beta$ are Chan-Paton factors describing
the $j^* \omega$-twisted bundles ${\cal E}|_T$ and ${\cal F}$,
the $\eta$ indices are tangent to $T$,
and (momentarily ignoring the twisting of \cite{abooetal,aboo2})
the $\theta$ indices are normal to $S \hookrightarrow X$.
Naively, the BRST cohomology classes of such states appear to be counted
by the sheaf cohomology groups
$H^n\left(T, {\cal E}^{\vee}|_T \otimes {\cal F} \otimes 
\Lambda^m {\cal N}_{S/X}|_T \right)$.

Exactly as in \cite{orig}, there is a spectral sequence
\begin{displaymath}
H^n\left(T, \left( {\cal E}|_T \right)^{\vee} \otimes {\cal F}
\otimes \Lambda^m {\cal N}_{S/X} \right) \: \Longrightarrow \:
\mbox{Ext}^{n+m}_{ X, \omega }\left( i_* {\cal E}, j_* {\cal F} \right)
\end{displaymath}
(See appendix~\ref{pfs} for proofs.)

Exactly as in \cite{orig}, we have neglected certain subtleties,
described in more detail in the previous section, and we conjecture
that when those subtleties are taken into account, we are left with
BRST cohomology classes counted by the $\omega$-twisted Ext groups above,
as we explicitly saw happens for parallel coincident branes in the
previous section.

\section{General intersecting branes} \label{gen}

The analysis for general intersections proceeds much as in the
above and in \cite{orig}.  Let $S$, $T$ be smooth submanifolds,
and let ${\cal E}$, ${\cal F}$ be twisted holomorphic vector
bundles on $S$, $T$, respectively, each twisted by the pullback
of an element of $H^2(X, {\cal O}_X^*)$.  In order to write down
the massless boundary Ramond sector states for the general case,
we need to take into account facts about branes at angles
\cite{bdl}.  We refer the reader to \cite{orig} for details
of the analysis in the untwisted case, and the analysis in the
twisted case is a nearly exact duplicate.
Suffice it to say that the massless boundary Ramond sector spectrum,
ignoring subtleties in boundary conditions, appears to be counted by the
sheaf cohomology groups
\begin{displaymath}
\begin{array}{c}
H^p\left( S \cap T, {\cal E}^{\vee}|_{S \cap T} \otimes
{\cal F}|_{S \cap T} \otimes \Lambda^{q-m} \tilde{N} \otimes
\Lambda^{top} {\cal N}_{S \cap T/T} \right) \\
H^p\left( S \cap T, {\cal E}|_{S \cap T} \otimes {\cal F}^{\vee}|_{
S \cap T} \otimes \Lambda^{q-n} \tilde{N} \otimes \Lambda^{top} {\cal N}_{S \cap T/S}
\right) 
\end{array}
\end{displaymath}
where
\begin{displaymath} 
\tilde{N} \: = \: TX|_{S \cap T} / \left( TS|_{S \cap T} + TT|_{S \cap T} \right)
\end{displaymath}
and $m = \mbox{rk } {\cal N}_{S \cap T/T}$, $n = \mbox{rk } {\cal N}_{S \cap T/S}$.

Just as in \cite{orig}, 
the line bundles $\Lambda^{top} {\cal N}_{S \cap T/T}$
and $\Lambda^{top} {\cal N}_{S \cap T/S}$ are a reflection of the
Freed-Witten anomaly \cite{freeded}.
Strictly speaking, the sheaf $i_* {\cal E}$ corresponds to a D-brane
with worldvolume bundle ${\cal E} \otimes \sqrt{ K_S^{\vee}}$,
to account for anomalies when the normal bundle is non-spin.
We have been able to ignore this subtlety previously in our discussion,
but here for general intersections it returns to haunt us.
When the open string B model is well-defined ({\it i.e.} taking into
account another anomaly), the ratios of the square roots of the
canonical bundles are the same as the line bundles above,
appearing in the spectral sequence.
The necessity of these line bundles can also be proven on purely
mathematical grounds, as a necessary requirement for finding a relationship
to Ext groups, and from Serre duality.
These subtleties are all discussed in great detail
in \cite{orig}, and we refer the reader there for more information.

Exactly as in \cite{orig}, there are two spectral sequences
\begin{eqnarray*}
\lefteqn{H^p\left( S \cap T, {\cal E}^{\vee}|_{S \cap T} \otimes
{\cal F}|_{S \cap T} \otimes \Lambda^{q-m} \tilde{N} \otimes
\Lambda^{top} {\cal N}_{S \cap T/T} \right) } \\
& \hspace{2.5in} \Longrightarrow & 
\mbox{Ext}^{p+q}_{ X, \omega }\left( i_* {\cal E}, j_* {\cal F} \right) \\
\lefteqn{H^p\left( S \cap T, {\cal E}|_{S \cap T} \otimes {\cal F}^{\vee}|_{
S \cap T} \otimes \Lambda^{q-n} \tilde{N} \otimes \Lambda^{top} {\cal N}_{S \cap T/S}
\right) } \\
& \hspace{2.5in} \Longrightarrow & \mbox{Ext}^{p+q}_{ X, \omega } \left( j_* {\cal F},
i_* {\cal E} \right)
\end{eqnarray*}
(See appendix~\ref{pfs} for proofs.)

All the subtleties that arose in \cite{orig} have immediate analogues
here.  For purposes of brevity, we merely refer the reader to \cite{orig}
for more details.  Just as in \cite{orig}, we conjecture that both of
the spectral sequences above are realized physically in BRST cohomology,
due ultimately to Chan-Paton-induced boundary condition twistings of
the form described in \cite{abooetal,aboo2}, as we explicitly saw happens
earlier in the special case of parallel coincident branes.

\section{Conclusions}

In this paper we have explicitly computed the massless boundary
Ramond sector spectrum in open strings between D-branes wrapped
on holomorphic submanifolds, with nontrivial flat $B$ fields
turned on, carrying on the program of \cite{orig,kps}.
Just as in our earlier work \cite{orig,kps}, we find that the
spectra are again counted by Ext groups, this time
Ext groups between twisted sheaves, where the twist is
determined by the $B$ field.

It has been previously proposed \cite{medc,mikedc,paulalb} that
some properties of D-branes could be modeled by derived categories.
The proposals in \cite{medc,mikedc,paulalb}, however, really
only speak to a special class of D-branes, namely, those
with no background fields turned on -- no $B$ field,
and no Higgs field vevs.  In this paper we begin to see how some
of those constraints can be lifted.
In the presence of a nontrivial flat $B$ field, for example,
we have given concrete evidence that ordinary derived categories
should be replaced by derived categories of twisted sheaves.
(See also \cite{kaporlov} for some previous work in this vein.)
It should also be possible to lift some of the other restrictions
mentioned.  For example, it has been suggested \cite{tomasme} that
giving nilpotent vevs to Higgs fields might be modeled
sheaf-theoretically in terms of sheaves on nonreduced subschemes
of Calabi-Yau's.  Work to compare open string spectra in
such backgrounds with Ext groups between sheaves on nonreduced 
subschemes is in progress.

\section{Acknowledgments}

We would like to thank D.~Morrison for useful conversations.  The work
of A.C.\ is partially supported by an NSF Postdoctoral Research
Fellowship.  The work of S.K.\ is partially supported by NSF grant DMS
02-96154 and NSA grant MDA904-02-1-0024.  The work of E.S.\ is
partially supported by NSF grant DMS 02-96154.

\appendix

\section{Proofs of spectral sequences}  \label{pfs}

In this appendix we derive the spectral sequences used in the rest of
the paper.  We work with a background smooth variety (or complex
manifold) $X$, endowed with a $B$-field $\alpha\in H^2(X, {\cal
O}_X^*)$.  We consider branes $(S, {\cal E})$ and $(T, {\cal F})$
supported at the smooth subvarieties $S$ and $T$ of $X$.  We let $W$ be
their scheme theoretic intersection, and we label the
inclusions as follows:
\begin{displaymath}
\xymatrix{
W \ar@{^{(}->}[r]^{k} \ar@{^{(}->}[d]^{l} & 
S \ar@{^{(}->}[d]_{i} \\
T \ar@{^{(}->}[r]^{j} & X.
}
\end{displaymath}
The bundles ${\cal E}$ and ${\cal F}$ are twisted by $i^*\alpha$ and $j^*\alpha$
on $S$ and $T$, respectively.  Note that although we do not assume
that $\alpha$ is torsion on $X$, the assumption that ${\cal E}$ and
${\cal F}$ exist (and have finite rank) implies that $i^*\alpha$ and $j^*\alpha$ are torsion
on $S$ and $T$, respectively.  We assume that $W$ is smooth; in
concrete terms this means that the set theoretic intersection of $S$
and $T$ is smooth, and furthermore the equality
\[ T_{w,W} = T_{w,S} \cap T_{w,T} \]
holds at every point $w\in W$. 

The main spectral sequence that we want to derive is:

\begin{theorem}
\label{app:mainthm}
Let $m$ be the rank of $N_{W/T}$, the normal bundle of $W$ in $T$, and
let $\tilde{N}$ be the vector bundle on $W$ given by
\[ \tilde{N} = T_X|_W / (T_S|_W + T_T|_W). \]
There exists a convergent spectral sequence
\[ {}^2E^{pq} = H^p(W, {\cal E}^{\vee}|_W \otimes {\cal F}|_W \otimes \Lambda^{m}
{\cal N}_{W/T} \otimes \Lambda^{q-m} \tilde{N}) \Longrightarrow \mbox{Ext}^{p+q}_X(i_*{\cal E},
j_*{\cal F}). \] 
We use the convention that $\Lambda^0 \tilde{N} = {\cal O}_W$ for a bundle $\tilde{N}$ on
$W$, {\em even if} $\tilde{N}=0$.  ${\cal E}^{\vee}$ denotes the dual of the bundle
${\cal E}$.
\end{theorem}

Since ${\cal E}$ and ${\cal F}$ are twisted by $i^*\alpha$ and $j^*\alpha$,
respectively, $i_*{\cal E}$ and $j_*{\cal F}$ are defined, and they are both
$\alpha$-twisted,~\cite[1.2.10]{andreithesis}.  Thus $\underline{\mbox{Ext}}^{p+q}_X(i_*{\cal E},
j_*{\cal F})$ is untwisted and hence $\mbox{Ext}^{p+q}_X(i_*{\cal E}, j_*{\cal F})$ makes
sense.  See~\cite[1.2.12]{andreithesis} for details.  Similarly, ${\cal E}^{\vee}|_W
\otimes{\cal F}|_W$ is untwisted, so the ${}^2E^{pq}$ term is defined.

Before we can proceed to the proof of Theorem~\ref{app:mainthm} we
need several auxiliary results.

\begin{proposition}
\label{app:prop1}
Around every point $w\in W$ there exists an open neighborhood $U
\subseteq X$ and a smooth subvariety $Y$ of $U$ containing $w$ such that
\begin{itemize}
\item[a)] $S\subseteq Y$, 
\item[b)] $W\cap U$ is the scheme theoretic intersection of $Y$ and $T$,
and
\item[c)] $Y$ and $T$ intersect in the expected dimension, i.e.
\[ \dim Y+\dim T-\dim X = \dim W. \]
\end{itemize}
\end{proposition}

\begin{proof}
Proceed by induction on $n = \dim T-\dim W$.  If $n=0$, then $T\subseteq
S$ and we can take $Y=X$.  If $n>0$, then $\dim W < \dim T$, and since we
assumed that
\[ T_{W,w} = T_{T,w}\cap T_{S,w}, \]
we conclude that $T_{T,w}$ is not contained in $T_{S,w}$.

Observe that $T_{S,w}$ is the intersection of the (Zariski) tangent
spaces $T_{X',w}$ of all the local hypersurfaces $X'$ through $w$ that
contain $S$. Since $T_{T,w}$ is not
contained in $T_{S,w}$, there exists a hypersurface $X'$ that contains
$S$ and such that $T_{T,w}\not\subseteq T_{X',w}$.  Thus $X'$ and $T'
= T\cap X$ are smooth at $w$, of dimensions $\dim X - 1$ and $\dim T
-1$, respectively, and $S\subseteq X'$.  We have
\[ S\cap T' = S\cap (X'\cap T) = (S\cap X')\cap T = S\cap T = W, \]
thus we are in the same situation as before, but with
$\dim T'-\dim W = n-1$.  By the induction hypothesis we can find $Y$
in $X'$ satisfying the conditions of the proposition.  Considering $Y$
inside of $X$ instead of $X'$ still satisfies (a) and (b), and we have
\begin{align*}
\dim Y+\dim T-\dim X & = \dim Y+(\dim T' + 1) - (\dim X'+1) \\
& = \dim Y+\dim T'-\dim X' = \dim W. 
\end{align*}
\end{proof}


\begin{proposition}
\label{app:prop2}
If $S\subseteq T$ then 
\[ \mbox{Tor}_q^X({\cal O}_S, {\cal O}_T) = \Lambda^q {\cal N}_{T/X}^{\vee}|_S, \]
where ${\cal N}_{T/X}$ is the normal bundle of $T$ in $X$.  (We abuse
notation and write ${\cal O}_S$ for $i_*{\cal O}_S$, etc.)
\end{proposition}

\begin{proof}
Since the question is local and $T$ is locally a complete
intersection, we can use the Koszul resolution with respect to the
normal bundle to obtain a locally free resolution of ${\cal O}_T$
\[ \cdots\ra \Lambda^{q+1} {\cal G}^{\vee} \ra \Lambda^q {\cal G}^{\vee} 
\ra \cdots \ra {\cal G}^{\vee} \ra {\cal O}_X \ra {\cal O}_T \ra 0 \] 
around every point of $T$, where ${\cal G}$ is locally free on $X$
and ${\cal G}|_T\simeq {\cal N}_{T/X}$.  
To compute $\mbox{Tor}_q^X({\cal O}_S, {\cal O}_T)$ we
tensor this resolution by ${\cal O}_S$, which has the effect of restricting
all the bundles to $S$ and making all the maps in the complex zero,
hence the result.
\end{proof}

\begin{proposition}[{\cite[Corollary to Th\'eor\`eme 4, p. V-20]{Ser}}]
\label{app:prop3}
If $W$ is of the expected dimension, i.e.,
\[ \dim S+\dim T - \dim X = \dim W, \]
then
\[ \mbox{Tor}_q^X({\cal O}_S, {\cal O}_T) = 0 \]
for $q>0$.
\end{proposition}

\begin{proposition}
\label{app:prop4}
\[ \mbox{Tor}_q^X({\cal O}_S, {\cal O}_T) = \Lambda^q \tilde{N}^{\vee}, \]
where $\tilde{N} = T_X/(T_S+T_T)$.
\end{proposition}

\begin{proof}
By Proposition~\ref{app:prop1}, around every point of $W$ we can find
$Y$ smooth such that $S$ is contained in $Y$, $T$ and $Y$ intersect
in the expected dimension, and $W=T\cap Y$.  

Since $T$ and $Y$ intersect in the expected dimension, using
Proposition~\ref{app:prop3} we see that the change-of-rings spectral
sequence~\cite[Exercise A3.45]{Eis}
\[ \mbox{Tor}_q^Y({\cal O}_S, \mbox{Tor}_p^X({\cal O}_T, {\cal O}_Y)) \Longrightarrow
\mbox{Tor}_{p+q}^X({\cal O}_S, {\cal O}_T) \]
consists of the single column
\[ \mbox{Tor}_q^Y({\cal O}_S, \mbox{Tor}_0^X({\cal O}_T, {\cal O}_Y)) = \mbox{Tor}_q^Y({\cal O}_S, {\cal O}_W) =
\Lambda^q {\cal N}_{S/Y}^{\vee}|_W, \]
where the last equality is Proposition~\ref{app:prop2}.  In other
words,
\[ \mbox{Tor}_q^X({\cal O}_S, {\cal O}_T) = \Lambda^q {\cal N}_{S/Y}^{\vee}|_W. \]

At each $w\in W$, the composition of maps
\[ {\cal N}_{S/Y,w} = T_{Y,w}/T_{S,w} \ra T_{X,w}/T_{S,w} \ra 
T_{X,w}/(T_{S,w}+T_{T,w}) = \tilde{N}_w \]
is injective, because a tangent vector to $Y$ will be tangent to $T$
only if it is tangent to $W$, and then it is tangent to $S$.  Comparing
dimensions, we see that the natural map ${\cal N}_{S/Y}\to
\tilde{N}$ is therefore an isomorphism.
Finally,
one can trace through the identifications and see that the local
isomorphism
\[ \mbox{Tor}_q^X({\cal O}_S, {\cal O}_T) \iso \Lambda^q \tilde{N}^{\vee} \]
does not depend on the choice of $Y$, hence it lifts to a global
isomorphism. 

As this calculation can be tedious, we supplement this argument with
an alternative approach.  The claimed result is elementary to check
directly for $q=0$.  For $q=1$ this is also straightforward: using the
exact sequence
\[
0\to \mathcal{I}_S\to \mathcal{O}_X\to \mathcal{O}_S\to 0,
\]
we compute that $\mbox{Tor}_1^X({\cal O}_S, {\cal O}_T)$ is isomorphic
to the kernel of the map
\[
\mathcal{I}_S\otimes\mathcal{O}_T\to \mathcal{O}_T,
\]
namely 
\[
\frac{\mathcal{I}_S\cap\mathcal{I}_T}{\mathcal{I}_S\mathcal{I}_T}.
\]
But this can be seen directly to be isomorphic to $\tilde{N}^{\vee}$; the
isomorphism is induced by the natural map
$(\mathcal{I}_S\cap\mathcal{I}_T)/(\mathcal{I}_S\mathcal{I}_T)\to 
\tilde{N}^{\vee}$ taking a class represented by a function
$f$ to the linear functional on $\tilde{N}$ induced
by $df$.

For $q>1$ we note the existence of a natural map
\begin{equation}
\label{higherq}
\Lambda^q\mbox{Tor}_1^X({\cal O}_S, {\cal O}_T)\to
\mbox{Tor}_q^X({\cal O}_S, {\cal O}_T).
\end{equation}
This alternating product is constructed by using the exterior algebra
structure on any local Koszul resolution and the multiplication on 
$\mathcal{O}_T$.  As the algebra structure on the Koszul complex is 
canonical (a change of regular sequence induces a canonical isomorphism
of algebras), the map (\ref{higherq}) is canonical, independent of the
choice of local defining equations yielding the Koszul complex.  Then
our earlier local calculations show that (\ref{higherq}) is an isomorphism.
\end{proof}

{}From here on we switch to the language of derived categories.
Recall~(\cite{andreithesis}, \cite{HarRD}) that objects in the derived category
of twisted coherent sheaves on a space (or abelian groups, etc.) are
complexes of such sheaves, with morphisms defined using the machinery
of derived categories.  For a morphism between smooth spaces (as well
as in much more general cases), one has derived functors corresponding
to push forward, pull back, $\otimes$, $\mbox{Hom}$, $\sHom$, etc.  Applying
any of these functors to a complex yields again a complex.  Given such
a complex ${\cal C}$ (of twisted sheaves, abelian groups, etc.) we can
consider its $q$-th cohomology object, denoted by ${\cal H}^q({\cal C})$.  It is
a twisted sheaf or an abelian group, accordingly.  We use the
convention that ${\cal H}_q({\cal C}) = {\cal H}^{-q}({\cal C})$.

Note that pushing forward by an affine morphism is exact, so we do not
need to derive this functor.  Also, tensoring by a locally free sheaf
is exact, and this tensoring operation does not need to be derived
either.  On the other hand, pulling back to a subvariety is only right
exact, so we need to left-derive it to get a well-defined functor on
derived categories.  Since resolutions by locally free sheaves of
finite rank may not exist (if $\alpha$ is not torsion on $X$, there
are no locally free sheaves of finite rank on $X$), we need to use resolutions
by ${\cal O}_X$-flat modules.  These are not quasi-coherent, but
resolutions using them exist for arbitrary
$\alpha$~(\cite[2.1.2]{andreithesis}), and since they are acyclic with
respect to the pull-back functor they can be used to compute derived
pull-backs by the general yoga of derived functors.

\begin{proposition}
\label{app:prop5}
We have
\[ {\cal H}_q(\Ld j^* i_* {\cal E}) = l_*({\cal E}|_W \otimes {\cal H}_q(\Ld i^* j_*
{\cal O}_T)|_W). \]
Combined with Proposition~\ref{app:prop4}, this yields
\[ {\cal H}_q(\Ld j^* i_* {\cal E}) = l_*({\cal E}|_W \otimes \Lambda^q \tilde{N}^{\vee}). \]
\end{proposition}

\begin{proof}
By the projection formula~(\cite[2.3.5]{andreithesis},~\cite[II.5.6]{HarRD}) we have
\[ j_* \Ld j^* i_* {\cal E} = i_*{\cal E} \lotimes_X j_* {\cal O}_T. \]
On the other hand, since $j_*$ is exact, we have
\[ {\cal H}_q (j_* \Ld j^* i_* {\cal E}) = j_* {\cal H}_q(\Ld j^* i_* {\cal E}), \]
and thus
\[ {\cal H}_q(\Ld j^* i_* {\cal E}) = j^* {\cal H}_q (j_* \Ld j^* i_* {\cal E}) = j^*
{\cal H}_q(i_*{\cal E} \lotimes_X j_* {\cal O}_T). \] 
(Here we have used the fact that for a sheaf or twisted sheaf, pulling
back {\em in the underived way} its push-forward by a closed embedding
yields back the original sheaf.)  Using the projection formula again
for $i_*$ we have
\[ i_* {\cal E} \lotimes_X j_* {\cal O}_T = i_* ({\cal E} \otimes_S \Ld i^* j_*
{\cal O}_T), \]
and since ${\cal E}$ is locally free on $S$ we get
\[ {\cal H}_q(i_* {\cal E} \lotimes_X j_* {\cal O}_T) = i_* {\cal H}_q({\cal E} \otimes_S \Ld i^* j_*
{\cal O}_T) = i_* ({\cal E} \otimes_S {\cal H}_q(\Ld i^* j_* {\cal O}_T)). \]
But for {\em sheaves} and {\em underived operations}, $j^* i_* = l_*
k^*$.  Therefore
\begin{eqnarray*}
{\cal H}_q(\Ld j^* i_* {\cal E}) & = & j^* i_* ({\cal E} \otimes_S {\cal H}_q(\Ld i^* j_*
{\cal O}_T))  \\
 & = & l_* k^* ({\cal E} \otimes_S {\cal H}_q(\Ld i^* j_* {\cal O}_T)) \\
 & = & l_*
({\cal E}|_W \otimes {\cal H}_q(\Ld i^* j_* {\cal O}_T)|_W), 
\end{eqnarray*}
which is the first result.  Applying again the projection formula we have
\[ {\cal H}_q(\Ld i^* j_* {\cal O}_T) = i^* {\cal H}_q(i_* {\cal O}_S \lotimes_X j_*
{\cal O}_T) = i^*\mbox{Tor}_q^X( {\cal O}_S, {\cal O}_T). \]
By Proposition~\ref{app:prop4}, 
\[ \mbox{Tor}_q^X({\cal O}_S, {\cal O}_T) = \Lambda^q \tilde{N}^{\vee}, \]
hence the second result.
\end{proof}

\begin{proposition}
\label{app:prop6}
If $l:W\hookrightarrow T$ is a locally complete intersection embedding
of varieties or complex analytic spaces of relative dimension $m$
(such is the case if, for example, W is a smooth subvariety of the
smooth variety $T$), and $\alpha$ is an element of $\mathrm{Br}(T)$, then we
have a natural isomorphism
\[ \R\sHom_T(l_*\cG, \cF) \iso l_* \R\sHom_W(\cG, l^!\cF) \]
for $\cG\in\D(W, l^*\alpha)$ and $\cF\in\D(T, \alpha)$, where
\[ l^!(\cF) = \Ld l^*(\cF) \otimes \Lambda^m {\cal N}_{W/T}[-m]. \]
(Here, ${\cal N}_{W/T} = (\cI_W/\cI_W^2)^{\vee}$, and $[-m]$ denotes the shift
by $-m$.)

Applying $\R\Gamma$ to both sides we conclude that $l^!$ is a right
adjoint to $l_*$, as functors between $\D(W, l^*\alpha)$ and $\D(T,
\alpha)$.
\end{proposition}

\begin{proof}
To prove the first statement it is enough to show that there exist
natural local isomorphisms 
\[ \R\sHom_U(l_*\cG|_U, \cF|_U) = l_* \R\sHom_V(\cG|_V, l^!(\cF|_V)) \]
for $U$ an open set of $T$, sufficiently small to trivialize $\alpha$,
and $V=U\cap W$.  (Here we work with either the analytic or the
\'etale topology, as the Zariski topology is not fine enough for the
Brauer group to be interesting.)  In this form, the above isomorphism
is nothing else than duality for a finite morphism~(\cite[III.7.3,
III.6.7 and Definition after III.1.3]{HarRD}; see~\cite{BanSta} for
similar statements for morphisms of analytic spaces).

The second statement follows immediately from the fact that 
\[ \R\Gamma(T, \R\sHom_T(\scdot, \scdot)) = \R\Hom_T(\scdot, \scdot)
\]
for $\alpha$-twisted sheaves~\cite[2.3.2]{andreithesis}.
\end{proof}

Now we can finally prove the main result.  \nopagebreak

{\it Proof of the main theorem.}
The functor $\Ld j^*$ is left adjoint to
$j_*$~(\cite[2.3.9]{andreithesis},~\cite[II.5.11]{HarRD}), so we have
\[ \Ext^{p+q}_X(i_*{\cal E}, j_* \cF) = {\cal H}^{p+q}\R\Hom_X(i_*{\cal E}, j_*\cF) =
{\cal H}^{p+q}\R\Hom_T(\Ld j^* i_* {\cal E}, {\cal F}). \] 
Writing down the Grothendieck spectral sequence for the composition of
the two functors $\R\Hom_T(\scdot, {\cal F})$ (contravariant) and $\Ld j^*
i_*(\scdot)$ (covariant) yields the convergent spectral sequence
\[ {}^2E^{pq} = {\cal H}^p\R\Hom_T({\cal H}_q(\Ld j^* i_* {\cal E}), {\cal F}) \Longrightarrow
{\cal H}^{p+q}\R\Hom_T(\Ld j^* i_* {\cal E}, {\cal F}). \]
(Note the implicit change of sign in $q$, due to the fact that
$\R\Hom_T(\scdot, \cF)$ is contravariant.)  Using
Proposition~\ref{app:prop5} we get that 
\[ {}^2E^{pq} = {\cal H}^p\R\Hom_T(l_*({\cal E}|_W \otimes \Lambda^q \tilde{N}^{\vee}), {\cal F}),
\]
which by Grothendieck duality (Proposition~\ref{app:prop6}) gives
\[ {}^2E^{pq} = {\cal H}^p\R\Hom_W({\cal E}|_W \otimes \Lambda^q \tilde{N}^{\vee}, {\cal F}|_W
\otimes \Lambda^{m} {\cal N}_{W/T}[-m]). \]
But ${\cal E}$ and $\Lambda^q \tilde{N}^{\vee}$ are locally free, so 
\begin{align*}
{}^2E^{pq} & = {\cal H}^p\R\Hom_W({\cal E}|_W \otimes \Lambda^q \tilde{N}^{\vee}, {\cal F}|_W
\otimes \Lambda^{m} {\cal N}_{W/T}[-m]) \\
& = H^{p-m}(W, {\cal E}^{\vee}|_W \otimes {\cal F}|_W \otimes \Lambda^{top} 
{\cal N}_{W/T} \otimes \Lambda^q \tilde{N}). 
\end{align*}
Renumbering and putting it all together we arrive at our result: there
exists a convergent spectral sequence
\[ {}^2E^{pq} = H^p(W, {\cal E}^{\vee}|_W \otimes {\cal F}|_W \otimes
\Lambda^{m} {\cal N}_{W/T} \otimes \Lambda^{q-m} \tilde{N}) \Longrightarrow
\Ext^{p+q}_X(i_*{\cal E}, j_* {\cal F}). \]

\hfill $\Box$

\section{Notes on the generalized Hodge conjecture} \label{ghc}

In this appendix we collect a few notes related to
the generalized Hodge conjecture.  In particular, when describing
the precise relationship between elements of $H^2(X, {\cal O}_X^*)$
and gerbes on Calabi-Yau threefolds, we encountered the group
\begin{displaymath}
H^3(X, {\bf Z}) \cap \left( H^{2,1}(X) \oplus H^{1,2}(X) \right)
\end{displaymath}
which plays an important role in the generalized Hodge conjecture in 
dimension three.   For more details on
this circle of ideas, see \cite{cg}.

Define the intermediate Jacobian
\[
J(X)=\left(H^{3,0}(X)\oplus H^{2,1}(X)\right)^*/\left(H^3(X,\mathbf{Z})
\right),
\]
a complex torus of dimension $h^{2,1}+1$.  Given a family of
holomorphic curves on the Calabi-Yau $X$, parametrized by $S$, there is an Abel-Jacobi
mapping
\[
\mathrm{Alb}(S)=\left(H^{1,0}(S)\right)^*/H_1(S)\to J(X).
\]
Now $J(X)$ is not an abelian variety, but it contains an
abelian variety $A(X)$ whose dimension is equal to the rank of
$H^3(X,{\bf Z})\cap (H^{2,1}(X)\oplus H^{1,2}(X))$.  In fact, $A(X)$ is the maximal
abelian variety contained in $J(X)$ and orthogonal to $H^{3,0}(X)$.
The generalized Hodge conjecture asserts that the Abel-Jacobi mapping is
surjective for some family of curves.

For example, on the Fermat quintic threefold, $H^3(X,{\bf Z})\cap
(H^{2,1}(X)\oplus H^{1,2}(X))$ has rank 200
and consequently $A(X)$ is an abelian variety 
of dimension 200.  In this case, it is straightforward to check that the
Abel-Jacobi image for the family of lines on $X$ is precisely $A(X)$.
Furthermore, $A(X)$ decomposes as a direct sum of genus 2 Jacobians.  This
genus 2 Jacobian has appeared in the physics literature in \cite{finite}
and the relevant mathematics has been discussed 
more recently in \cite{mustatathesis}.


\newpage

\begin{thebibliography}{199}

\addcontentsline{toc}{section}{References}


\bibitem{paulron} P. Aspinwall and R. Donagi,
``The heterotic string, the tangent bundle, and derived categories,''
Adv. Theor. Math. Phys. {\bf 2} (1998) 1041-1074, 
{\tt hep-th/9806094}.

\bibitem{medc} E. Sharpe, ``D-branes, derived categories, and
Grothendieck groups,'' Nucl. Phys. {\bf B561} (1999) 433-450,
{\tt hep-th/9902116}.

\bibitem{mikedc} M. Douglas, ``D-branes, categories, and 
${\cal N}=1$ supersymmetry,'' J. Math. Phys. {\bf 42} (2001) 2818-2843,
{\tt hep-th/0011017}.

\bibitem{paulalb} P. Aspinwall and A. Lawrence, ``Derived categories
and zero-brane stability,'' JHEP 0108 (2001) 004, {\tt hep-th/0104147}.



\bibitem{orig} S. Katz and E. Sharpe, ``D-branes, open string vertex
operators, and Ext groups,''
{\tt hep-th/0208104}.


\bibitem{kps} S. Katz, T. Pantev, and E. Sharpe, ``D-branes, orbifolds,
and Ext groups,'' {\tt hep-th/0212218}.


\bibitem{hitchin} N. Hitchin, ``Lectures on special Lagrangian submanifolds,''
{\tt math.DG/9907034}.


\bibitem{medt} E. Sharpe, ``Discrete torsion,'' {\tt hep-th/0008154}.

\bibitem{dtrev} E. Sharpe, ``Recent developments in discrete torsion,''
Phys. Lett. {\bf B498} (2001) 104-110, {\tt hep-th/0008191}.

\bibitem{dougdt} M. Douglas, ``D-branes and discrete torsion,'' 
{\tt hep-th/9807235}.

\bibitem{kap} A. Kapustin, ``D-branes in a topologically nontrivial
$B$-field,'' {\tt hep-th/9909089}.

\bibitem{bm} P. Bouwknegt and V. Mathai, ``D-branes, $B$ fields, and
twisted K-theory,'' JHEP 0003 (2000) 007,
{\tt hep-th/0002023}.


\bibitem{andreithesis} A. C\u ald\u araru, 
``Derived categories of twisted sheaves
on Calabi-Yau manifolds,'' Cornell University Ph.D. thesis, 2000.
Available at {\tt http://www.math.upenn.edu/\~{}andreic}


\bibitem{freeded} D. Freed and E. Witten, ``Anomalies in string theory
with D-branes,'' {\tt hep-th/9907189}.




\bibitem{eh} D. Eisenbud and J. Harris, {\it The Geometry of Schemes},
Springer-Verlag, New York, 2000.

\bibitem{kaporlov} A. Kapustin and D. Orlov, ``Vertex algebras,
mirror symmetry, and D-branes:  the case of complex tori,''
{\tt hep-th/0010293}.



\bibitem{wzw1} C. Klimcik and P. Severa, ``Open strings and D-branes in
WZNW models,'' Nucl. Phys. {\bf B488} (1997) 653-676,
{\tt hep-th/9609112}.

\bibitem{wzw2} A. Alekseev and V. Schomerus, ``D-branes in the WZW model,''
Phys. Rev. {\bf D60} (1999) 061901,
{\tt hep-th/9812193}.

\bibitem{agm} P. Aspinwall, D. Morrison, and M. Gross, ``Stable singularities
in string theory,'' Comm. Math. Phys. {\bf 178} (1996) 115-134,
{\tt hep-th/9503208}.



\bibitem{edtft}  E. Witten, ``Mirror manifolds and topological
field theory,'' in {\it Mirror Symmetry I}, ed. S.-T. Yau,
American Mathematical Society, 1998.


\bibitem{abooetal} A. Abouelsaood, C. Callan, C. Nappi, and S. Yost,
``Open strings in background gauge fields,''
Nucl. Phys. {\bf B280} (1987) 599-624.

\bibitem{aboo2} C. Callan, C. Lovelace, C. Nappi, and S. Yost,
``String loop corrections to beta functions,''
Nucl. Phys. {\bf B288} (1987) 525-550.


\bibitem{bdl} M. Berkooz, M. Douglas, and R. Leigh, ``Branes intersecting
at angles,'' Nucl. Phys. {\bf B480} (1996) 265-278, {\tt hep-th/9606139}.


\bibitem{tomasme} T. Gomez and E. Sharpe, ``D-branes and scheme theory,''
{\tt hep-th/0008150}.


\bibitem{Ser}
J.-P. Serre, {\em Alg\`ebre Locale - Multiplicit\'es}, LNM 11,
  Springer-Verlag, 1965.


\bibitem{Eis}
D. Eisenbud,  {\em Commutative algebra with a view toward algebraic
geometry}, LNM 150, Springer-Verlag, 1995.


\bibitem{HarRD}
R. Hartshorne, {\em Residues and Duality}, LNM 20, Springer-Verlag, 1966.




\bibitem{BanSta} C. B\u anic\u a and   O. St\u an\u a\c sil\u a, 
  {\em Algebraic methods in the global theory of complex spaces}, John Wiley,
  New York, 1976.



\bibitem{cg} C.H.\ Clemens and P.A.\ Griffiths, ``The intermediate
Jacobian of the cubic threefold,'' Ann.\ Math.\ {\bf 95} (1972), 281--356.

\bibitem{finite} P. Candelas, X. de la Ossa, and F. Rodriguez-Villegas,
``Calabi-Yau manifolds over finite fields, I,'' {\tt hep-th/0012233}.

\bibitem{mustatathesis} A. Musta\c t\u a, Ph.D. thesis, University of Utah, 2003.







\end{thebibliography}
\end{document}